\documentclass{jpsj3}

\usepackage{graphicx,amsmath,amssymb,bm}

\allowdisplaybreaks 

\usepackage{color}
\definecolor{Green}{rgb}{0,0.7,0}
\newcommand{\e}{ {\rm e}}
\newcommand{\ET}{ $\alpha$-(BEDT-TTF)$_2$I$_3$}

\newcommand{\bk}{ \bm{k}}
\newcommand{\ef}{ E_{\rm F}}
\newcommand{\eD}{ E_{\rm D}}
\newcommand{\bkD}{ \bm{k}_{\rm D}}

\newcommand{\ka}{k_x}
\newcommand{\kb}{k_y}
\newcommand{\kc}{k_z}

\newcommand{\kabc}{k_x+k_y+k_z}
\newcommand{\kab}{k_x+k_y}
\newcommand{\kbc}{k_y+k_z}
\newcommand{\kac}{k_x+k_z}

\begin{document}


\title{
Novel Dirac Electron in Single-Component Molecular Conductor 
[Pd(dddt)$_2$] (dddt=5,6-dihydro-1,4-dithiin-2,3-dithiolate)  
}
\author{
Reizo Kato$^{1}$ and 
Yoshikazu Suzumura$^{2}$
\thanks{E-mail: suzumura@s.phys.nagoya-u.ac.jp}
}
\inst{
$^1$
RIKEN, 2-1 Hirosawa, Wako-shi, Saitama 351-0198, Japan \\
$^2$
Department of Physics, Nagoya University,  Chikusa-ku, Nagoya 464-8602, Japan \\
}

\hspace{8 cm}

\abst{
 Dirac electrons in a single-component molecular  conductor [Pd(dddt)$_2$]  (dddt=5,6-dihydro-1,4-dithiin-2,3-dithiolate)  under  pressure  have been examined  using a tight-binding model which 
 consists of  highest occupied molecular orbital (HOMO) and  lowest unoccupied molecular orbital (LUMO) functions in four molecules per unit cell.
The Dirac cone between the conduction and valence bands originates from  the property  that the HOMO has ungerade symmetry and the LUMO has gerade symmetry.
 The Dirac point forms a loop in the three-dimensional Brillouin zone, 
 which is symmetric with respect to the plane of $k_y=0$, 
where $k_y$ is the intralayer momentum along the molecular stacking direction,  i.e., with the largest (HOMO-HOMO, LUMO-LUMO) transfer energy.   
 The parity at  time reversal invariant momentum (TRIM) is calculated  using the inversion symmetry around the lattice point of the crystal. 
 It is shown that  such an exotic Dirac electron can be  understood from the parity of the wave function  at the TRIM and also from an effective Hamiltonian. 
}


\maketitle

\section{Introduction} 
  Since the discovery of the quantum Hall effect in graphene,\cite{Novoselov2005_Nature438}  
 two-dimensional (2D) massless Dirac fermions have been a fascinating topic.
In addition to the graphene with  monolayer, 
    a Dirac electron was found in organic conductor \ET$\;$ as  a bulk system,   \cite{Katayama2006_JPSJ75} and  
 the properties of molecular Dirac fermion systems have been 
 studied extensively. 
\cite{Kajita_JPSJ2014} 
 
\begin{figure}
  \centering
\includegraphics[width=12cm]{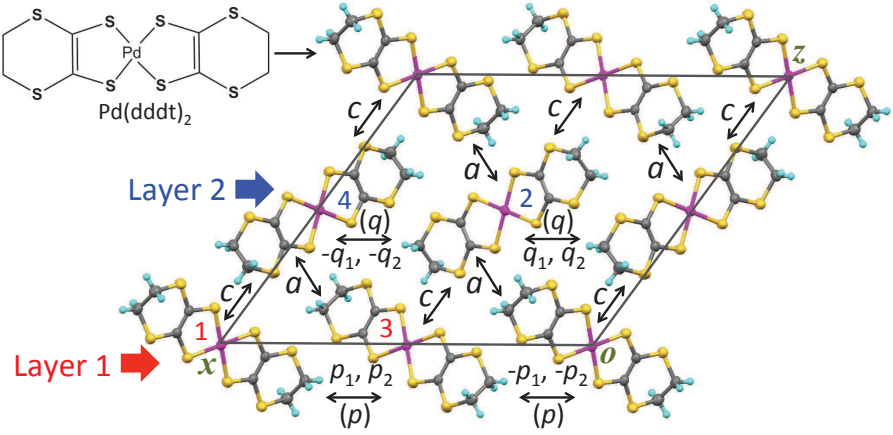}   
  \caption{
 (Color online) 
Crystal structure of [Pd(dddt)$_2$] viewed along the $y$ axis, 
 with four molecules (1, 2, 3, and 4) per unit cell, 
 two of which (molecules 1 and 2) are crystallographically independent. 
 Each [Pd(dddt)$_2$] molecule is on the inversion center. 
 The coordinates of the Pd atoms for molecules 1, 2, 3, and 4 are
  (1,0,0), (1/2,1/2, 1/2), (1/2,1/2,0), and (1,0,1/2), respectively. 
 There are two kinds of layers, layer 1 (molecules 1 and 3) 
 and layer 2 (molecules 2 and 4), 
 each of which consists of crystallographically equivalent molecules. 
[Pd(dddt)$_2$] molecules are uniformly stacked along the $y$ axis.
}
\label{fig:structure}
\end{figure}

Recently, a Dirac electron was found in the single-component molecular conductor [Pd(dddt)$_2$] 
 (dddt=5,6-dihydro-1,4-dithiin-2,3-dithiolate), 
 which shows  a constant resistivity with decreasing temperature under pressure.
\cite{Kato_JACS,Cui}
Based on  first-principles calculation, which shows the existence of a Dirac cone,
\cite{Tsumuraya}
a tight-binding model of [Pd(dddt)$_2$] 
 consisting  of 
 highest occupied molecular orbital (HOMO) and  lowest unoccupied molecular orbital (LUMO)  functions in  four molecules per unit cell 
 was  proposed.\cite{Kato_PSJ_2015} 
In the crystal, there are two crystallographically independent layers 
given by  layers 1 and 2 (Fig.~\ref{fig:structure}), and  the Dirac cone 
 originates from 
 the   HOMO-based band in layer 1  and the LUMO-based band in layer 2.
\cite{Kato_JACS} 
The interplay of the interlayer and intralayer transfer integrals is crucial to  obtain the Dirac point.\cite{Kato_JACS}
 The existence of such a Dirac point is clarified as follows
 for a simple case of 2D momentum  
 with a fixed interlayer momentum.  
 Since the difference in the energy level between the  HOMO and LUMO is  small,
  the HOMO band is located higher  than the LUMO band around the $\Gamma$ point, 
  and then the overlap  between them results in a Fermi line
     in the absence of the HOMO-LUMO transfer energies. 
 However the Fermi line disappears in the presence of the HOMO-LUMO
 transfer energies owing to the opening of a gap, which comes from 
 the  combined  effect of the intralayer HOMO-LUMO transfer energies 
  and  the interlayer HOMO-HOMO/LUMO-LUMO transfer energies.
Furthermore, there is a line (nodal line)  passing through the $\Gamma$ point, 
 on which  the HOMO-LUMO coupling  vanishes. 
Thus, the  Dirac point 
 is obtained at the intersection of the Fermi line and  the  nodal line 
 owing to the closing of the gap.
\cite{Kato_JACS} 
Such a nodal line originates from a property 
 of the HOMO-LUMO transfer energy. 
The [Pd(dddt)$_2$] molecule has an inversion center at the Pd atom. 
For the HOMO with a bonding $\pi$ character, 
the inversion produces a phase change for the molecular orbital. 
 This means that the HOMO has ungerade (odd) symmetry. 
On the other hand, the LUMO with an antibonding $\pi$ character 
has gerade (even) symmetry, where the inversion results in the same phase 
 for the orbital. 
 Due to these properties, the HOMO-LUMO transfer energies  interrelated by 
 the center of symmetry have opposite signs in the [Pd(dddt)$_2$] crystal.

 In addition to the Dirac cone on the  $k_x$-$k_y$ plane of the intralayer 
 2D momentum, 
  a novel feature of [Pd(dddt)$_2$] is 
 the formation of a loop of the Dirac point with varying  
 the interlayer momentum $k_z$
      in the three-dimensional (3D) Brillouin zone.
\cite{Kato_PSJ_2016_march} 
  For the case of \ET$\:$,  the  negligible interlayer  
 HOMO-HOMO  transfer integral 
    suggests  a  Dirac line that is open 
 at the boundary of the first Brillouin zone.\cite{Katayama2008_JPSJ}
The Dirac electron in [Pd(dddt)$_2$]  is exotic  since the Dirac point 
 originates  from the interplay of the HOMO and LUMO functions. 
The purpose of the present paper is to clarify such Dirac electron by examining    the  Dirac point in the 3D Brillouin zone. 
 The result is analyzed  by calculating  the parity of the wave function 
 at   the TRIM (time reversal invariant momentum),
  which comes from the inversion symmetry around the molecular site. 
\cite{Suzumura_PSJ_2016}
In Sect. 2, the model and formulation are given.
In Sect. 3, Dirac points under a pressure of 8 GPa corresponding to a previous
 experiment  are calculated as a function of $\bm{k}= (k_x, k_y, k_z)$, where  $k_y$ 
 is  the intralayer momentum 
 along the molecular stacking direction, i.e., 
with the largest transfer energy. 
 It is shown that  a pair of Dirac points forms a loop 
 in the 3D Brillouin zone. 
In Sect. 4, the mechanism for the formation of  the loop of the Dirac point 
   is analyzed in terms of  the parity  at the TRIM,
 which  is 
 calculated for both the pressure of 8 GPa and the ambient pressure to comprehend
  the emergence of the Dirac point.
A summary and discussion in terms of the effective Hamiltonian are given 
 in Sect. 5.

\section{Model and Formulation}
\subsection{Tight-binding model}
 The crystal structure of [Pd(dddt)$_2$] is shown in 
  Fig.~\ref{fig:structure},
 which consists of four molecules (1, 2, 3, and 4)
  with HOMO and LUMO functions in the unit cell.
The crystal structure was 
 determined by the single-crystal X-ray diffraction 
 method at ambient pressure ($P$=0) and was estimated by first-principles 
 density functional theory (DFT)
  calculations at $P$= 8 GPa.\cite{Kato_JACS}    
Transfer energies are given by pairs between nearest-neighbor molecules, 
  where those between the layers are  expressed  as   
  $a$ (molecules 1 and 2 and molecules 3 and 4), and  $c$ (molecules 1 and 4  
 and molecules 2 and 3),  those in the same layer are expressed as 
 $p$ (molecules 1 and 3) and $q$ (molecules 2 and 4), 
  and  those along the stacking axis are  given by 
    $b$ (Fig.~\ref{fig:structure}).  
Based on the crystal structure, 
 we examine  the tight-binding model Hamiltonian  given by 
\begin{equation}
H = \sum_{i,j} t_{i,j;\alpha, \beta} |i, \alpha> <j, \beta| \; ,
\label{eq:H_model}
\end{equation}
where  $i$ and $j$ are the  sites  of the unit cell 
  with total number $N$,
     and $\alpha$ and $\beta$ denote the eight molecular orbitals 
      given by the HOMO $(H1, H2, H3, H4)$ and 
         LUMO $(L1, L2, L3, L4)$.  
 The lattice constant is taken as unity. 
The transfer energies, $t_{i,j;\alpha, \beta}$,  are classified 
 as 
   HOMO-HOMO (HH), LUMO-LUMO (LL),  and HOMO-LUMO (HL) transfer energies. 
Taking eV as the unit of energy,  
the transfer energies $t_{i,j;\alpha, \beta}$ 
 under a pressure of  $P$ = 8 GPa\cite{Kato_JACS}  (0 GPa) 
are given by 
$a_{H}=-0.0345 \;(-0.0136) $, 
$a_{L}=-0.0 \;(-0.0049) $,
$a_{HL}=0.0260 \;(0.0104 ) $,
$b_{1H}=0.2040 \;( 0.112 ) $,
$b_{1L}=0.0648 \;(0.0198) $,
$b_{1HL}=0.0219 \;( 0.0214) $,
$b_{2H}=0.0762 \;(0.0647 ) $,
$b_{2L}=-0.0413 \;(0.0) $,
$b_{2HL}=-0.0531 \;(-0.0219 )$,
$c_{H}=0.0118 \;(0.0 )$,
$c_{L}=-0.0167 \;(-0.0031)$,
$c_{HL}=0.0218 \;(0.0040)$,
$p_{H}=0.0398 \;(0.0102 ) $,
$p_{L}=0.0205 \;(0.0049 )$,
$p_{1HL}=-0.0275 \;(-0.0067 )$,
$p_{2HL}=-0.0293 \;(-0.0074 )$,
$q_{H}=0.0247 \;(0.0067 )$,
$q_{L}=0.0148 \;(0.0037 )$,
$q_{1HL}=-0.0186 \;(-0.0048 )$, and 
$q_{2HL}=-0.0191 \;(-0.0051 )$. 
The transfer energies were calculated using  
 the extended  H\"uckel method. 
The gap between the energy of the HOMO and that of the LUMO is taken 
 as $\Delta E = $ 0.696 eV to reproduce 
  the energy band of the first-principles calculation.

Using the Fourier transform 
$ |\alpha(\bm{k})>= \sum_{j} \exp[- i \bm{k}\bm{r}_j] \; |j,\alpha>$
       with  wave vector  $\bk = (k_x, k_y, k_z)$, 
Eq.~(\ref{eq:H_model}) is calculated as\cite{Kato_JACS}
\begin{equation}
H = 
 \sum_{\bm{k}} |\Phi(\bm{k})> \hat{H}(\bm{k}) <\Phi(\bm{k})|\; , 
\label{eq:H} 
\end{equation}
 where $\hat{H}(\bm{k})$ is  the Hermite  matrix Hamiltonian 
 with the matrix elements $t_{\alpha,\beta}$ defined by 
\begin{eqnarray} 
 t_{\alpha,\beta} & = & \left( \hat{H}(\bk) \right)_{\alpha,\beta}  
 \; ,
\label{eq:H_m}
\end{eqnarray} 
 and the base is given by 
$<\Phi(\bm{k})| = (<H1|,<H2|,<H3|,<H4|, <L1|, <L2|, <L3|, <L4|).$ 
The matrix elements of $t_{\alpha,\beta}$  
are given  in the Appendix. 
Since the symmetry of the HOMO (LUMO) 
 is odd (even)  with respect to the 
 Pd atom, the matrix element of H-L (H-H and L-L) 
 is the odd (even) function 
 with respect to $\bm{k}$. 
The energy band  $E_j(\bk)$ 
 and the wave function $\Psi_j(\bk)$, $(j = 1, 2, \cdots, 8)$ 
 are calculated from 
\begin{equation}
\hat{H}(\bm{k}) \Psi_j(\bk) 
 = E_j(\bk) \Psi_j(\bk) \; , 
\label{eq:energy_band}
\end{equation}
 where $E_1 > E_2 > \cdots > E_8$ and 
\begin{equation}
\Psi_j(\bm{k}) = \sum_{\alpha}
 d_{j,\alpha}(\bk) |\alpha> \; ,
\label{eq:wave_function}
\end{equation}
 with $\alpha =$  
 H1, H2, H3, H4, L1, L2, L3, and L4. 
Noting that the band is  half-filled owing to the HOMO and LUMO 
functions,
 we examine the gap defined by 
\begin{equation}
E_g(\bk) = {\rm min} (E_4(\bk)-E_5(\bk)) \; ,
\label{eq:Eg}
\end{equation}
for all $\bk$ in the Brillouin zone. 
The Dirac point $\bk_D$ is obtained from $E_g (\bkD)= 0$.

\subsection{Parity at TRIM}
In order to analyze the  Dirac point, 
  we calculate the parity at the TRIM 
 given by $\bm{G}$/2 with $\bm{G}$ being the reciprocal lattice vector, 
 where 
  $\bm{G}/2 =(0,0,0)$, $(\pi, 0 , 0)$,  $ (0,\pi,0)$, and  $(\pi,\pi,0)$
 correspond to the $\Gamma$, X, Y, and M points,   
and  $\bm{G}/2 =(0, 0, \pi)$, $(\pi, 0, \pi)$,  $ (0,\pi,\pi)$, and  $(\pi, \pi , \pi)$
 correspond to the 
  Z, D, C, and E points, respectively. 
  Applying the case of  \ET $\;$ with 
the 4 $\times$ 4 matrix Hamiltonian\cite{Piechon2013} 
 to the present case of the 8  $\times$ 8 matrix Hamiltonian, 
 the inversion with respect to 
   molecular site  1 in Fig.~\ref{fig:structure} gives 
the matrix for  the  translation of the base,  $\hat{P}(\bm{G}/2)$, 
  expressed as  
\begin{eqnarray} 
 \hat{P}(\bk) & =  &
\begin{pmatrix}
 -1 & 0 & 0 & 0 &            0 & 0 & 0 & 0 \\ 
  0 & -{\rm e}^{ -i k_x - i k_y - ik_z} & 0 & 0 &            0 & 0 & 0 & 0 \\ 
  0 & 0 & -{\rm e}^{ -i k_x -i k_y } & 0 &            0 & 0 & 0 & 0           \\ 
  0 & 0 & 0 & -{\rm e}^{ i k_z } &            0 & 0 & 0 & 0  \\ 
  0 & 0 & 0 & 0 &         1 & 0 & 0 & 0 \\ 
  0 & 0 & 0 & 0 &            0 &  {\rm e}^{ -i k_x - i k_y - ik_z}  & 0 & 0 \\ 
  0 & 0 & 0 & 0 &            0 & 0 &  {\rm e}^{ -i k_x -i k_y } & 0 \\ 
  0 & 0 & 0 & 0 &            0 & 0 & 0 &  {\rm e}^{ i k_z }  \\ 
\end{pmatrix} .
 \nonumber \\
\label{eq:P}
\end{eqnarray} 
The relation  
$(\hat{P}(\bm{k}))_{{\rm Hj},{\rm Hj}}
= - (\hat{P}(\bm{k}))_{{\rm Lj},{\rm Lj}}$
for $j$=1, 2, 3,and 4 
 originates  from the fact that the HOMO has ungerade symmetry and the LUMO has gerade symmetry.
The eigenvalue and eigenfunction 
 ($\alpha$ = H1, H2, $\cdots$, L4)
 are obtained from  
\begin{equation}
\hat{P}(\bm{k}) u_\alpha 
 = p_\alpha(\bk) u_\alpha \; , 
\label{eq:parity_eigen}
\end{equation}
where 
 $p_\alpha =  (\hat{P}(\bm{k}))_{\alpha,\alpha }$, 
   $u_{\rm H1}(\bk)^t = (1,0,0,0,0,0,0,0)$, 
  $u_{\rm H2}^t = (0,1,0,0,0,0,0,0)$, 
 $\cdots$, and 
  $u_{\rm L4}^t = (0,0,0,0,0,0,0,1)$. 
At the TRIM,  one obtains  $p_\alpha(\bm{G}/2) = + (-)$,  which gives the 
 parity. 
For example, the odd parity  $p_\alpha(\bm{G}/2) = -$  
 at  the TRIM  is  given by 
  $\alpha$ = H1, H2, H3, H4 for the $\Gamma$ point,   
  $\alpha$ = H1, H4, L2, L3  for the X point, 
  $\alpha$ = H1, H4, L2, L3  for the Y point, 
  $\alpha$ = H1, H2, H3, H4  for the M point, 
  $\alpha$ = H1, H3, L2, L4 for the Z point,  
  $\alpha$ = H1, H2, L3, L4 for the D point, 
  $\alpha$ = H1, H2, L3, L4 for the C point, and 
  $\alpha$ = H1, H3, L2, L4 for the E point.  
Note that  the wave function at the  $\Gamma$ point 
with odd parity is  given only by the  HOMO 
 and that  at the Z point is given by  both the  LUMO and HOMO.

Since $[\hat{P}(\bm{G}/2), \hat{H}(\bm{G}/2)] = 0$, 
   $\Psi_j(\bm{G}/2)$ is also an eigenfunction of 
 $\hat{P}(\bm{G}/2)$.
 Then,  at the TRIM, we obtain the parity from  
\begin{eqnarray}
 \label{eq:parity_a}
 \hat{P}(\bm{G}/2) \Psi_j(\bm{G}/2) = P_{Ej}(\bm{G}/2) \Psi_j(\bm{G}/2) 
 \;  
\end{eqnarray}
 with  $P_{Ej}(\bm{G}/2) = + (-)$,  
 which denotes the even (odd) parity. 
 Note that 
$d_{j \alpha}$ in Eq.~(\ref{eq:wave_function}) vanishes 
 for $\alpha$ when the parity $p_\alpha(\bm{G}/2)$  is  opposite to 
 that of $P_{Ej}(\bm{G}/2)$.
Using Eq.~(\ref{eq:wave_function}), 
Eq.~(\ref{eq:parity_a})  is rewritten as 
\begin{eqnarray}
 P_{Ej}(\bm{G}/2) = \Psi_j(\bm{G}/2)^\dagger \hat{P}(\bm{G}/2 ) \Psi_j(\bm{G}/2) = \sum_{\alpha = {\rm H1}}^{\rm L4} p_{\alpha}(\bm{G}/2) |d_{j \alpha}|^2 \; .
 \label{eq:parity_b}
\end{eqnarray}

 In Eq.~(\ref{eq:parity_a}), we used the notation
  $P_{Ej}(\Gamma)$, $P_{Ej}(X)$, $P_{Ej}(Y)$, $P_{Ej}(M)$, 
  $P_{Ej}(Z)$, $P_{E_j}(D)$, $P_{Ej}(C)$, and $P_{Ej}(E)$, 
   for $\bm{k}$ corresponding to 
  the $\Gamma$, X, Y, M, Z, D, C, and E  points, respectively. 
We  define  $P_{\delta}$ [
 = $P(k_z=0)$,  $P(k_y=0)$, and $P(k_z = \pi)$]
  as 
\begin{subequations}
 \label{eq:2D_parity}
\begin{eqnarray} 
P(k_z=0) & = & \prod_{j=5}^{8}P_{E_j}(\Gamma)P_{E_j}(X) P_{E_j}(Y)P_{E_j}(M) \; ,
\label{eq:2D_parity_a}   
 \\
 P(k_y=0)
   &=& \prod_{j=5}^{8}P_{E_j}(Z)P_{E_j}(\Gamma) P_{E_j}(X)P_{E_j}(D) \; ,
 \label{eq:2D_parity_b} \\
 P(k_z = \pi)
   &=& \prod_{j=5}^{8}P_{E_j}(Z)P_{E_j}(D) P_{E_j}(C)P_{E_j}(E) \; ,
 \label{eq:2D_parity_c} 
\end{eqnarray}
\end{subequations}
where  each $P_{\delta}$ denotes a quantity  assigned on a plane including 
the four respective   TRIMs. 
The condition for the Dirac point  between $E_4$ and $E_5$ 
 is given by\cite{Fu2007_PRB76,Piechon2013} 
\begin{equation}
 P_{\delta} = -1 \; , \; +1  \; . 
\label{eq:Dirac_cond}
\end{equation}
 When  $P_{\delta} = -1 (+1)$, the number of pairs of Dirac points  
  between $E_4$ and $E_5$ is odd (zero or even).
This fact can be understood from the idea of the $\pi$  jump for the $Z_2$ Berry phase.
\cite{Kariyado2013_PRB} 
Note that Eqs.~
(\ref{eq:2D_parity_a}),
(\ref{eq:2D_parity_b}), and 
(\ref{eq:2D_parity_c})
 describe  the condition of the Dirac point  on the planes of $k_c=0$ (TRIM with the $\Gamma$, X, Y, and M points) 
 and  $k_b=0$   (TRIM with the Z, $\Gamma$, X, and D points),
and  $k_c=\pi$ (TRIM with the Z, D, C, and  E points), respectively.

\section{Loop of Dirac Point in Three Dimensions}
\begin{figure}
  \centering
\includegraphics[width=6cm]{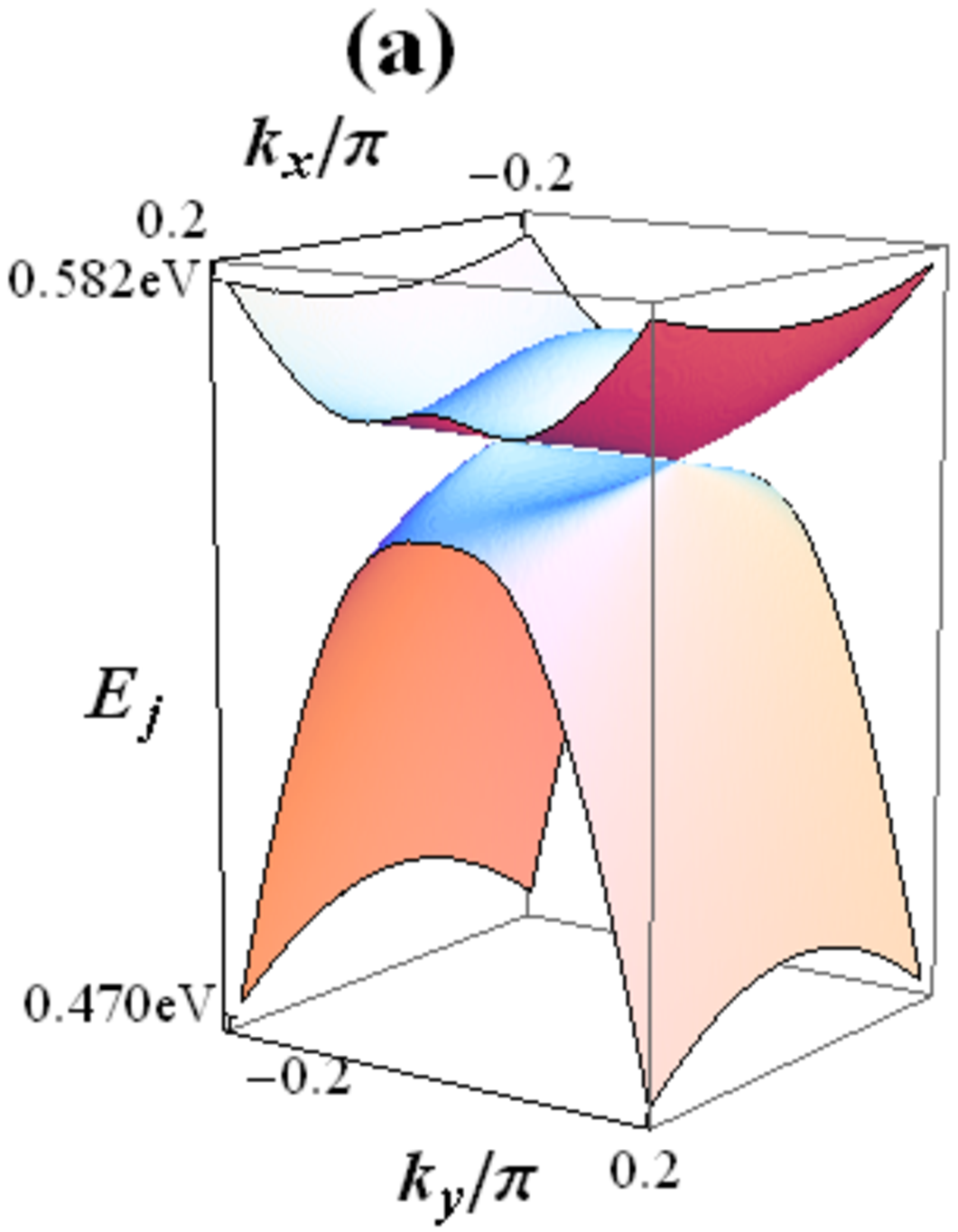}  
\includegraphics[width=6cm]{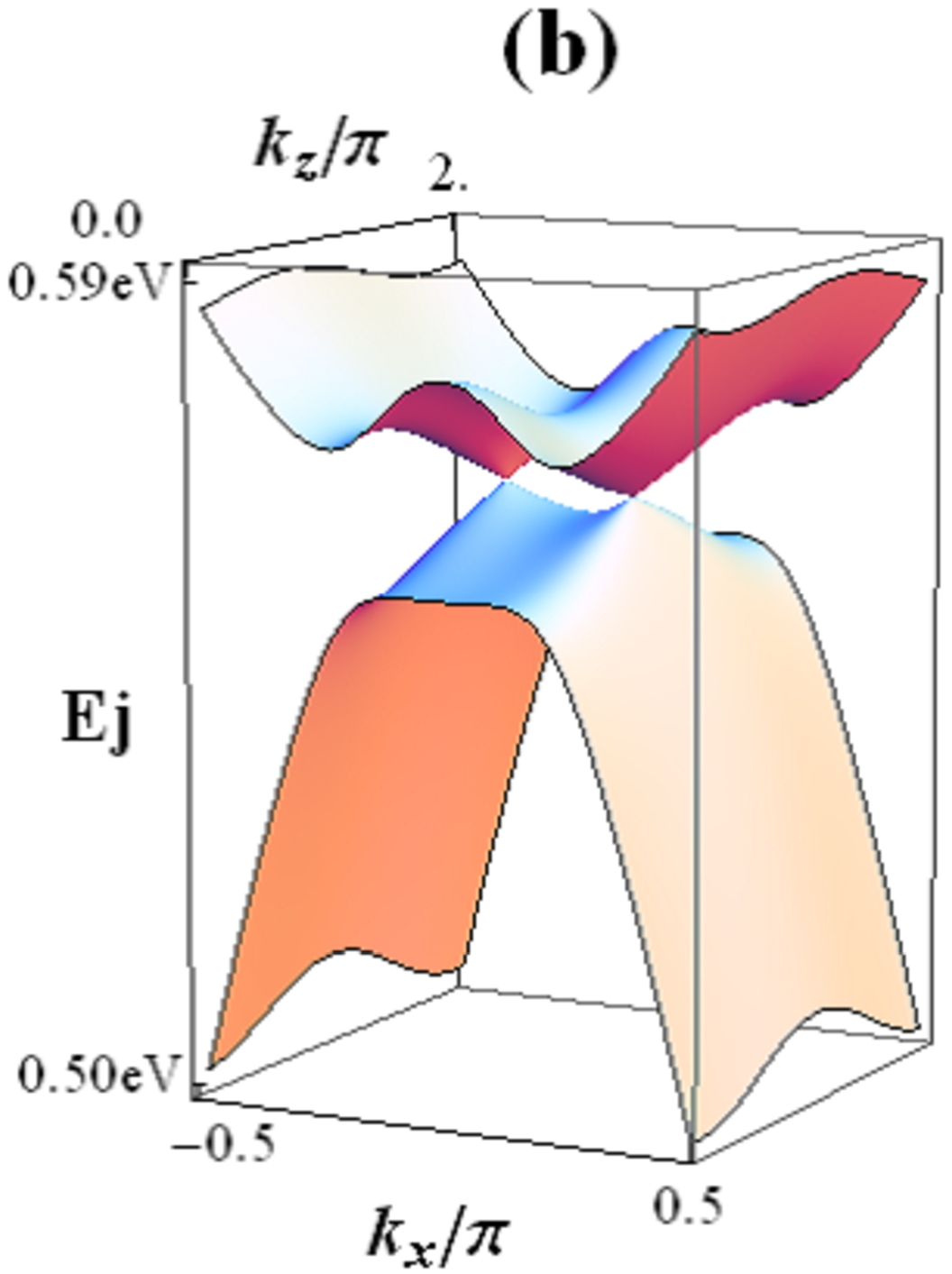}  
\caption{(Color online)
  Energy bands of $E_4(\bm{k})$ and $E_5(\bm{k})$ 
      with  fixed $k_z=0$ (a) and $k_y=0$ (b),
  where the Dirac point is given by  
 $\bkD / \pi =(0, \pm 0.0875 ,0)$  and  
$( \pm 0.155, 0, \pm 1.09)$, respectively. 
}
\label{fig2}
\end{figure}

\begin{figure}
  \centering
\includegraphics[width=4cm]{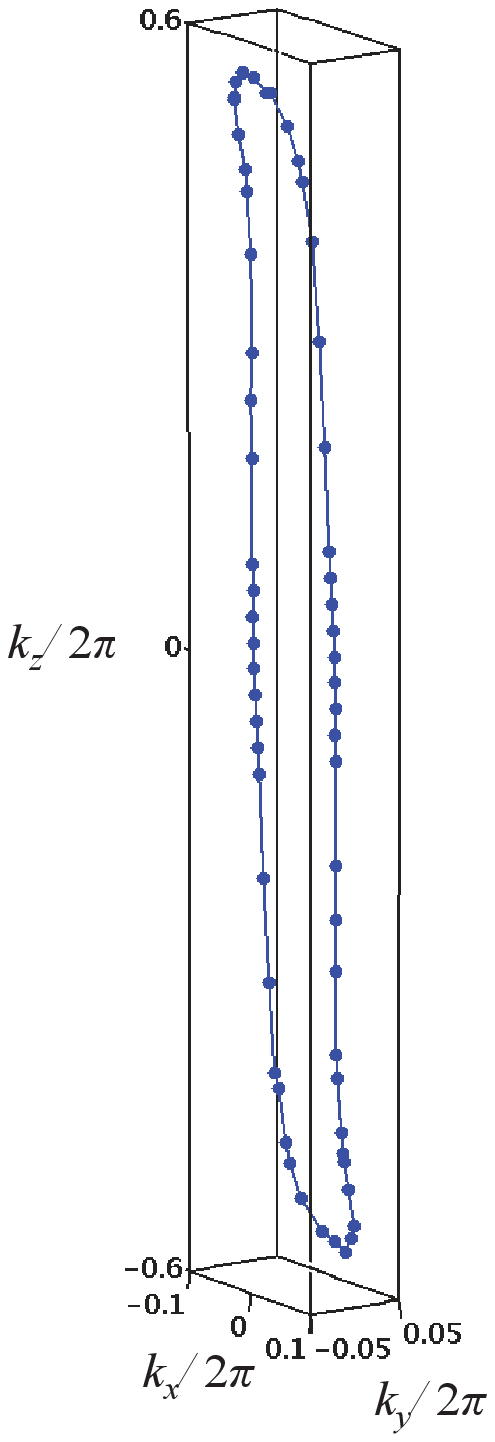} \\  
\includegraphics[width=6cm]{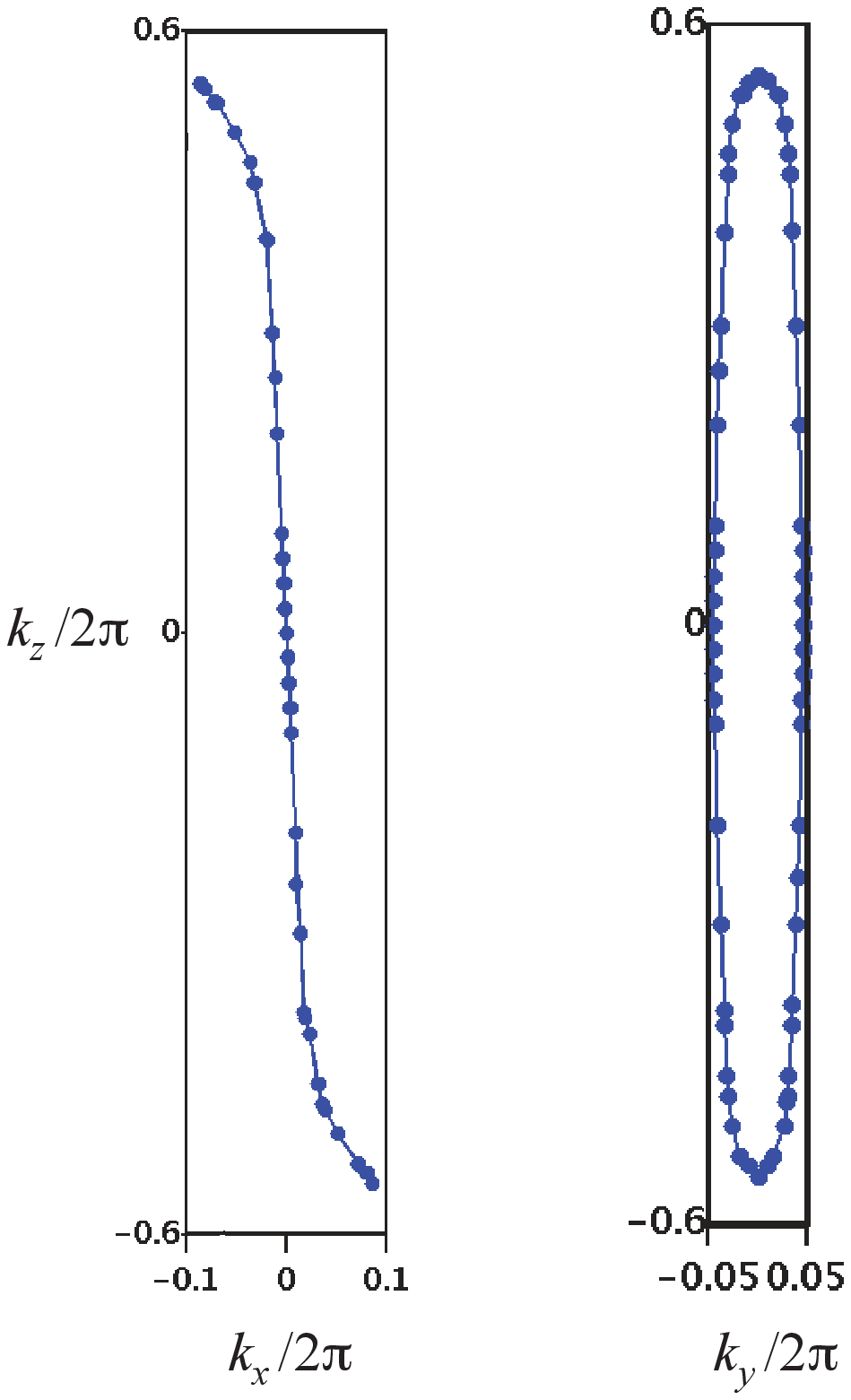}   
\caption{
(Color online) 
 Loop (upper panel) formed by  Dirac points  in the 3D 
extended Brillouin zone.  The lower panel denotes the loop projected 
 on  the $k_x$-$k_z$ and $k_y$-$k_z$ planes. 
}
\label{fig3}
\end{figure}

First we examine the energy bands $E_4(\bk)$ and $E_5(\bk)$ 
  with the Dirac point for two typical cases of $k_z$ = 0 and $k_y$ = 0, 
   which are shown on the $k_x$-$k_y$ plane in Fig.~\ref{fig2}(a) 
  and on the $k_x$-$k_z$ plane in Fig.~\ref{fig2}(b), respectively. 
A Dirac point with the tilted Dirac cone exists between 
 $E_4(\bk)$ and $E_5(\bk)$,  where the upper (lower) band denotes 
 $E_4(\bk)$ ($E_5(\bk)$) corresponding to the conduction (valence) band. 
In Fig.~\ref{fig2}(a), the center denotes the $\Gamma$ point 
 ($\bk$ = (0, 0, 0)) and the Dirac points are located on the line of 
 $k_x$ = 0. The convex upward region in $E_4(\bk)$ is mainly determined 
 by the HOMO function in layer 1, 
 while the  convex downward region in $E_5(\bk)$ 
   is mainly determined by the LUMO function in layer 2. 
 The energy of $E_4(\bk)$ ($E_5(\bk)$) in the remaining region originates 
  from the LUMO (HOMO) function. 
  This means that the HOMO-based band in layer 1 and the LUMO-based band 
 in layer 2  cross around the $\Gamma$ point.\cite{Kato_JACS}  
In Fig.~\ref{fig2}(b), 
  the center denotes the Z point ($\bk$ = (0, 0, $\pi$)) 
 and the Dirac points are disposed symmetrically with respect to the Z point. 
 The saddle point of $E_4(\bk)$ can be  seen at the Z point. 
 The tilted Dirac cone is elongated along the $k_z$ axis in Fig.~\ref{fig2}(a)
 and along the $k_y$ axis in Fig.~\ref{fig2}(b). 
 This suggests that the Dirac cone is overturned with increasing $k_z$.

Next we examine the Dirac point in the 3D wave vector space. 
Since  significant interlayer (along the $k_z$ direction) 
 transfer integrals exist, the system has a 3D character. 
 In particular, the HOMO-LUMO couplings that play a crucial role 
 in the Dirac cone formation exhibit $k_z$ dependence.\cite{Kato_JACS}  
Figure \ref{fig3}  demonstrates that the Dirac point moves in a manner 
 depending on $k_z$ and describes a loop in the extended Brillouin zone. 
The loop is symmetrical with respect to the plane of $k_y$ = 0, 
 while the symmetry of the Dirac point with respect to the $\Gamma$ point 
 is due to the time-reversal symmetry of the Hamiltonian. 
With increasing $|k_z|$, a crossover from the Dirac cone 
 on the $k_x$-$k_y$ plane to that on the $k_z$-$k_x$ plane occurs, 
 for example, the Dirac cone is already elongated along the $k_y$ axis 
 for the Dirac points 
 $\bkD/(2\pi)$ = $\pm(-0.0425, \pm 0.0265, 0.25)$. 
The loop is almost parallel to the $k_y$-$k_z$ plane around $k_z/2\pi$ = 0, 
 while the loop is turned up from the $k_y$-$k_z$ plane 
 with increasing $|k_z/2\pi|$ up to 0.545 (Fig.~\ref{fig3}).
Regarding the number of  Dirac points on the $k_x$-$k_y$ 
 plane with a fixed $k_z$ value in the reduced Brillouin zone, 
 there are two Dirac points for 
 $0 <k_z/\pi < 0.91$ and $1.09 < k_z/\pi< 2.0$, 
 and four Dirac points for $0.91 < k_z/\pi < 1.09$. 
 In the next section, we show another aspect of two pairs of Dirac points 
 at $k_z = \pi$.

 Here we note a slight variation of the energy on the loop, 
 which gives an electron pocket around $k_z$ = 0 (Fig.~\ref{fig2}(a)) 
 and a hole pocket around $k_y$ = 0 (Fig.~\ref{fig2}(b)), 
 implying a nodal line semi-metal.\cite{Murakami2007,Burkov2011} 
In fact, 
 the energy at the Dirac point in Fig.~\ref{fig2}(a) (Fig.~\ref{fig2}(b)) 
  is $\simeq  0.002$ V ($\simeq 0.001$ eV) below (above) 
the Fermi energy\cite{Suzumura_HPSP17}.
\begin{figure}
  \centering
\includegraphics[width=6cm]{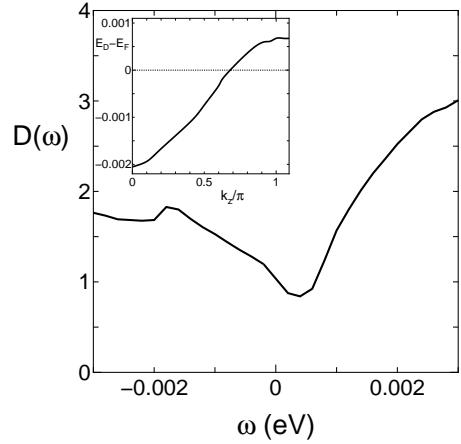}  
\caption{
Density of states $D(\omega)$ in the unit of (eV)$^{-1}$ 
 as a function of $\omega$ (eV), where $\omega=0$ corresponds to 
the Fermi energy $\ef$=0.5561 eV. The inset denotes $E_D-\ef$ 
 as a function of $k_z/\pi$, where $E_{\rm D}$ 
 is the energy at the Dirac point $\bkD$.
}
\label{fig4}
\end{figure}

In order to understand  the electronic properties associated with 
 such a loop of the Dirac point, we examine  the density of states (DOS)
  per spin and per site, which  
 is given by\cite{Suzumura_HPSP17}  
\begin{eqnarray}
D(\omega) &=& \frac{1}{N} \sum_{kz} 
\left[ \sum_{kx, ky} \sum_{j}
 \delta (\omega - E_j(\bk)) \right] \; ,
  \label{eq:dos}
\end{eqnarray}
where the Fermi energy, $\ef (=0.5561)$,  is obtained 
from $\int_{-\infty}^{\ef} d \omega D(\omega)$ = 2 
due to a half-filled band. 
The quantity $[\cdots]$ corresponds to the 2D  DOS for the fixed $k_z$, which is 
 proportional to $|\omega - \eD|$
 with $\eD (= E_4(\bkD) = E_5(\bkD))$  close to the Dirac point. 
As shown in the inset of Fig.~\ref{fig4},
 $\eD$   exhibits a  monotonic increase 
 with increasing  $k_z/\pi$  from  0 to 1.09  
 due to the  $k_z$ dependence of $\bkD$.
The total DOS obtained after the summation of $k_z$ 
 is shown in Fig.~\ref{fig4}, 
 where $\omega$ is measured from $\ef$. 
This variation of  $\eD$ results in $D(0) \not= 0$, 
 i.e.,  metallic behavior for small $\omega$.
The location of the dip due to the Dirac cone is slightly above $\ef$ 
 and the width of the dip is smaller than  that of $\eD$
 since the number of Dirac points above $\ef$ is larger than that 
 below $\ef$ as seen from Fig.~\ref{fig3}. 
 This asymmetry of the DOS may give  rise to a characteristic  
 temperature ($T$) dependences of the  magnetic susceptibility\cite{Suzumura_HPSP17}
 and specific heat compared with those of the 2D Dirac cone, 
 which are  proportional  to $T$ and $T^2$, respectively.

\section{Dirac Point vs. Parity at TRIM}
 Now we examine the Dirac points using the 
 condition Eq.~(\ref{eq:Dirac_cond}).   
Since there is  a degeneracy between $E_1$ and $E_2$ ($E_3$ and $E_4$ )
   at certain  TRIMs, 
      a  small amount of site potential is added to the diagonal elements 
        in $\hat{H}(\bm{k})$  to obtain the parity of Eq.~(\ref{eq:parity_b}).
In the present case,  
  we add a  potential  $- 0.0001$ eV to  
 $t_{H3,H3}$,  $t_{H4,H4}$, 
 $t_{L3,L3}$, and $t_{L4,L4}$ 
    in Eq.~(\ref{eq:H_m}).
Some of the  $P_{Ej}(\bm{G}/2)$ depend on 
    the choice of such potential but  
         Eq.~(\ref{eq:Dirac_cond}) remains the same.
From Eq.~(\ref{eq:parity_b}), we obtain  64 parities, where 
  half of them correspond to those of the filled band.
Note that  $\Pi_{j=1}^8 P_{Ej}(\bm{G}/2)$ = + since  half of the eight 
eigenvalues of Eq.~(\ref{eq:P}) are negative  for every $\bm{G}/2$. 
   The respective parities $ P_{Ej}(\bm{G}/2)$ at $P$ = 8 GPa 
 are listed in Table \ref{table_1}. 

\begin{table}
\caption{ 
 Parities  $ P_{Ej}(\bm{G}/2) (= \pm) $ at $P$ = 8 GPa, 
  where the TRIMs are 
  $\bm{G}/2 =(0,0,0)$ ($\Gamma$), $(\pi, 0 , 0)$ (X), 
   $ (0,\pi,0)$ (Y),  $(\pi,\pi,0)$ (M),
   $(0, 0, \pi)$ (Z), $(\pi, 0, \pi)$ (D),
   $ (0,\pi,\pi)$ (C), and  $(\pi, \pi , \pi)$ (E).
 }
\begin{center}
\begin{tabular}{cccccccccc}
\hline\noalign{\smallskip}
   &  $E_1$ & $E_2$ & $E_3$ & $E_4$ 
                                  &  $E_5$ & $E_6$ & $E_7$ & $E_8$  \\
\noalign{\smallskip}\hline\noalign{\smallskip}
$ P_{Ej}(\Gamma)$   & $+$  & $+$  & $+$  & $-$     & $+$  & $-$  & $-$  & $-$   \\
$ P_{Ej}({\rm X})$  & $+$  & $-$  & $-$  & $+$     & $-$  & $+$  & $+$  & $-$   \\
$ P_{Ej}({\rm Y})$  & $-$  & $+$  & $+$  & $-$     & $+$  & $-$  & $-$  & $+$   \\
$ P_{Ej}({\rm M})$  & $+$  & $+$  & $+$  & $+$     & $-$  & $-$  & $-$  & $-$   \\
$ P_{Ej}({\rm Z})$  & $+$  & $+$  & $-$  & $-$     & $-$  & $+$  & $-$  & $+$   \\
$ P_{Ej}({\rm D})$  & $+$  & $-$  & $+$  & $-$     & $-$  & $+$  & $-$  & $+$   \\
$ P_{Ej}({\rm C})$  & $+$  & $-$  & $+$  & $-$     & $-$  & $+$  & $-$  & $+$   \\
$ P_{Ej}({\rm E})$  & $-$  & $-$  & $+$  & $+$     & $+$  & $+$  & $-$  & $-$   \\
\\
\noalign{\smallskip}\hline
\end{tabular}
\end{center}
\label{table_1}
\end{table}

Substituting the parities of Table \ref{table_1} into  
Eqs.~(\ref{eq:2D_parity_a}), (\ref{eq:2D_parity_b}),  and (\ref{eq:2D_parity_c}), we obtain  
\begin{subequations}
 \label{eq:2D_parity}
\begin{eqnarray} 
 P(k_z=0) & = & -1
 \; ,\label{eq:2D_parity_8a}   
 \\
  P(k_y=0)    &=& -1 
 \; , \label{eq:2D_parity_8b} \\
   P(k_z = \pi)  &=& +1 
 \; . \label{eq:2D_parity_8c} 
\end{eqnarray}
\end{subequations}
Equations (\ref{eq:2D_parity_8a}) and  (\ref{eq:2D_parity_8b})
 are consistent 
 with the fact that  
 there is a pair of Dirac points for $k_z=0$ 
 and for $k_y=0$  as shown in Figs.~\ref{fig2}(a) and 2(b), respectively. 
 The loop of  the  Dirac point 
in  the  Brillouin zone 
 can be understood from  Eqs.~(\ref{eq:2D_parity_8a}) and (\ref{eq:2D_parity_8b}) 
  as follows.
Equation (\ref{eq:2D_parity_8a}) shows a pair of  Dirac points located on  the planes of $k_z=0$,  while  Eq.~(\ref{eq:2D_parity_8b}) displays another pair 
 of  Dirac points on the plane of $k_y=0$ being perpendicular to that of 
 $k_z=0$. 
Noting that the lines connecting these Dirac points are symmetric with respect   to $k_y=0$, the existence of the Dirac point on the plane of $k_y=0$ 
 suggests the loop of the Dirac point. 
Equation (\ref{eq:2D_parity_8c}) is also consistent with the fact that 
 two pairs of Dirac points exist for $k_z=\pi$ as mentioned in the previous section.

Here we numerically examine  the difference in the parity between  $P_{Ej}(\Gamma)$  and $P_{Ej}({\rm Z})$  with $j$= 4 and 5, which gives 
 Eqs.~(\ref{eq:2D_parity_8a}) and (\ref{eq:2D_parity_8c}). 
 The components of the wave function 
 $d_{j,\alpha}(\bk)$ in Eq.~(\ref{eq:wave_function}) 
 are estimated  as 
\begin{subequations}
\begin{eqnarray}
 \label{eq:wave_component_a}
& & (|d_{j,H_1}|^2, |d_{j,H_2}|^2,|d_{j,H_3}|^2,|d_{j,H_4}|^2,
|d_{j,L,1}|^2, |d_{j,L,2}|^2, |d_{j,L,3}|^2, |d_{j,L,4}|^2) \nonumber \\
\;\;\; &\simeq&  
( 0.490,0.010, 0.490, 0.010, 0, 0, 0, 0)\;,  \;\;
      {\rm for} \;\; E_4(\Gamma),  \\
 \label{eq:wave_component_b}
\;\;\; &\simeq&  
(0, 0, 0, 0, 0.014, 0.486, 0.014, 0.486) 
\;, \;\; {\rm for} \;\;  E_5(\Gamma), \\
 \label{eq:wave_component_c}
\;\;\; &=&  
( 0.5, 0, 0.5, 0, 0, 0, 0, 0) 
\;, \;\; {\rm for} \;\; E_4({\rm Z}),  \\  
 \label{eq:wave_component_d}
\;\;\; & = &  
 (0, 0, 0, 0, 0, 0.5, 0, 0.5 ) 
\;,\;\; {\rm for} \;\;  E_5({\rm Z}) .   
\end{eqnarray}
\end{subequations} 
 Substituting Eqs.~(\ref{eq:wave_component_a})-(\ref{eq:wave_component_d})
  into Eq.~(\ref{eq:parity_b}),
 we obtain 
 $ P_{E4}(\Gamma)= -$, $ P_{E5}(\Gamma) = +$, 
   $ P_{E4}({\rm Z}) = - $, and $ P_{E5}({\rm Z}) = -$, respectively, 
 where 
 $(p_{\rm H1}(\bm{G}/2), p_{\rm H2}(\bm{G}/2), p_{\rm H3}(\bm{G}/2), 
   p_{\rm H3}(\bm{G}/2),$  $p_{\rm L1}(\bm{G}/2),  p_{\rm L2}(\bm{G}/2),
 p_{\rm L3}(\bm{G}/2), p_{\rm L4}(\bm{G}/2))$ 
 = $(-,-,-,-,+,+,+,+)$ for the $\Gamma$ point  
   and  $(-,+,-,+,+,-,+,-)$ for the Z point  from Eq.~(\ref{eq:P}).
Both $E_4(\Gamma)$ and $E_4({\rm Z})$ 
 are determined by  the HOMO function while 
 both $E_5(\Gamma)$ and $E_5({\rm Z})$ 
 are determined by  the LUMO  function.
The difference in the parity between $E_4$ and $E_5$ at the $\Gamma$ point 
is understood  from  the LUMO and  HOMO  functions,
 whose  parities are  different from each other. 
However, at the Z point, the parity of $E_4$ is the same as that of $E_5$, 
  although the former (latter) is described by  the HOMO (LUMO) 
 function. 
Actually, the same parity of $P_{E4}({\rm Z})=P_{E5}({\rm Z})= -$  
 is obtained  from Eqs.~(\ref{eq:wave_component_c}) 
and (\ref{eq:wave_component_d}) with
$ p_{\rm H1}({\rm Z})=p_{\rm H3}({\rm Z})
 =  p_{\rm L2}({\rm Z})=p_{\rm L4}({\rm Z})= -$. 
We also see,  from Eq.~(\ref{eq:P}),
 that the product of the parities of 
 $\Psi_4$(Z), $\Psi_6$(Z), $\Psi_7$(Z), and $\Psi_8$(Z) 
  becomes positive  since 
 all of these wave functions are given by the HOMO function.
Thus, we obtain $P(k_z=\pi)=+1$ 
owing to $P_{E4}({\rm Z})=P_{E5}({\rm Z})= -$, 
 although $\Psi_5$(Z) is given by the LUMO function. 
 Furthermore we note that the loop of the Dirac point 
is  associated with the fact  
  that  $\Psi_{5}$(Z) is determined by $L2$ and $L4$, i.e.,
     the  LUMO function is determined by layer 2. 

\begin{table}
\caption{ 
The parity of  $ P_{Ej}(\Gamma)$ and $ P_{Ej}({\rm Z})$
 for $P$=0    where 
 those of $ P_{Ej}(\bm{G}/2)$ with X ,Y, M ,D, C and E are the same as 
  Table \ref{table_1}. 
 }
\begin{center}
\begin{tabular}{cccccccccc}
\hline\noalign{\smallskip}
$P=0$        & $E_1$              & $E_2$ & $E_3$ & $E_4$ 
                                  &  $E_5$ & $E_6$ & $E_7$ & $E_8$  \\
\noalign{\smallskip}\hline\noalign{\smallskip}
$ P_{Ej}(\Gamma)$   & $+$  & $+$  & $+$  & $+$     & $-$  & $-$  & $-$  & $-$   \\
$ P_{Ej}({\rm Z})$  & $+$  & $+$  & $-$  & $-$     & $-$  & $-$  & $+$  & $+$   \\
\noalign{\smallskip}\hline
\end{tabular}
\end{center}
\label{table_2}
\end{table}

Finally, we note the emergence  of the Dirac point where 
 a pair of Dirac points appears at one of the TRIMs.\cite{Montambaux} 
In Table \ref{table_2}, the parities at the $\Gamma$ point and the Z point 
 at $P$=0 (ambient pressure) are shown to compare with those at $P$=8 GPa 
 in Table \ref{table_1}.
It turns out that  the Dirac point is absent at $P$ = 0 owing to 
 $P(k_z=0)$= + (i.e., $P_{E4}(\Gamma)$= + and $P_{E5}(\Gamma) = -$ ). 
Using a linear interpolation for  the transfer energy 
 between $P=0$ and 8 GPa,  we obtain that  the emergence of the loop of the Dirac point occurs at $P$ $\simeq$ 7.6 GPa.  
   For the $\Gamma$ point, the level crossing between $E_4(\Gamma)$ and 
 $E_5(\Gamma)$ followed by the emergence of a pair of  Dirac points 
   occurs owing to the different parity.
For the Z point, the level crossing between $E_4$(Z) and $E_5$(Z) 
also occurs for $P \simeq$ 7.8 GPa owing to the different 
functions  of the 
    HOMO and LUMO  but with  the same parity, resulting in  
 a loop within the first Brillouin zone for 
 $7.6 \; {\rm GPa} < P < 7.8 \; {\rm GPa}$.

\section{Summary and Discussion}
We examined the Dirac point in  the single-component molecular conductor [Pd(dddt)$_2$]   
 within a tight-binding model that  consists of  
 HOMO and LUMO functions in four molecules in the unit cell. 
 It is crucial for the present Dirac electron 
 that the HOMO has the ungerade symmetry and the LUMO has the gerade symmetry. 
 We obtained a loop of the Dirac point 
 in the 3D Brillouin zone as a result of the 
 combined effect of the interlayer HOMO-HOMO/LUMO-LUMO and intralayer 
 HOMO-LUMO couplings. 
The present Dirac point  is exotic since a conventional molecular 
 Dirac electron  system  with only a single  molecular orbital
 gives  a line of the Dirac point 
 extending in the Brillouin zone as shown in \ET. 
 From the calculation of  the Dirac point
 for both  fixed $k_z$ and  fixed $k_y$, we found that
 the plane  displaying the Dirac cone rotates with increasing $k_z$, 
 which  comes from the combined effect  of the interlayer and intralayer 
 matrix elements.

The Dirac point was analyzed using  the  parity  of the wave function 
 at the TRIM. The parity was calculated from the matrix of Eq.~(\ref{eq:P}), 
 which comes from  the inversion symmetry with respect to 
 the lattice site of  the Pd atom   
  and  describes 
 the difference in the symmetry between the HOMO and the LUMO.  
The behavior of the parity well explains the loop and the emergence of the Dirac point.  
The conditions of the Dirac point given by 
Eqs.~(\ref{eq:2D_parity_8a}), (\ref{eq:2D_parity_8b}), 
 and (\ref{eq:2D_parity_8c}) 
 support  the existence of a  loop of the Dirac point. 
The  parities of $E_4$ and $E_5$ are different for the $\Gamma$ point
 while they are  the same for the Z point.   
This difference in  the parity between the $\Gamma$ and Z points 
   is compatible with  the respective behaviors of the Dirac point 
at the $\Gamma$ and Z points, 
   where, for a fixed $k_z$, 
 a pair of Dirac points exists  for the former case and 
  two pairs of Dirac points  exist  for the latter case.

\begin{figure}
  \centering
\includegraphics[width=6cm]{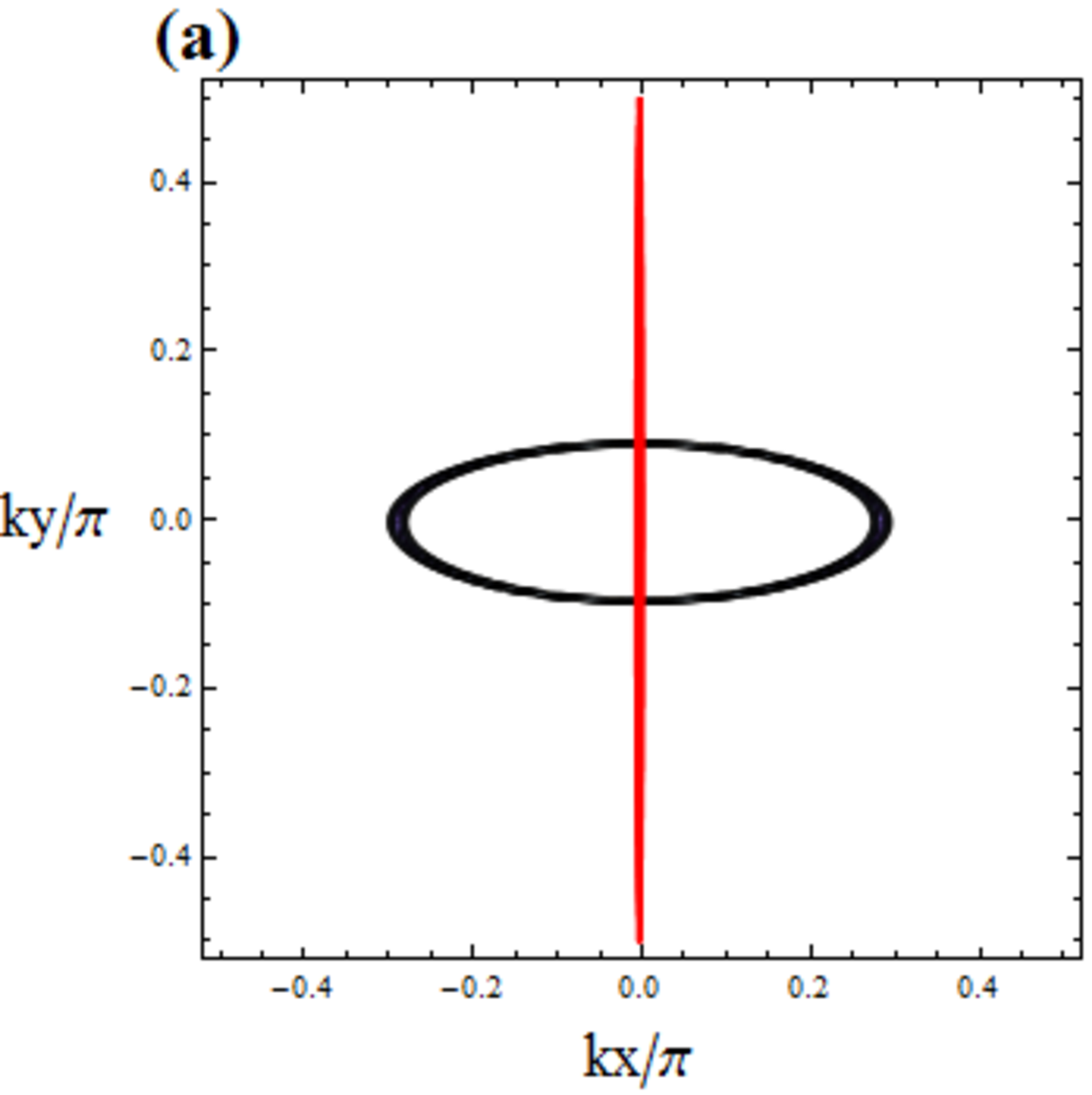}  
\includegraphics[width=6cm]{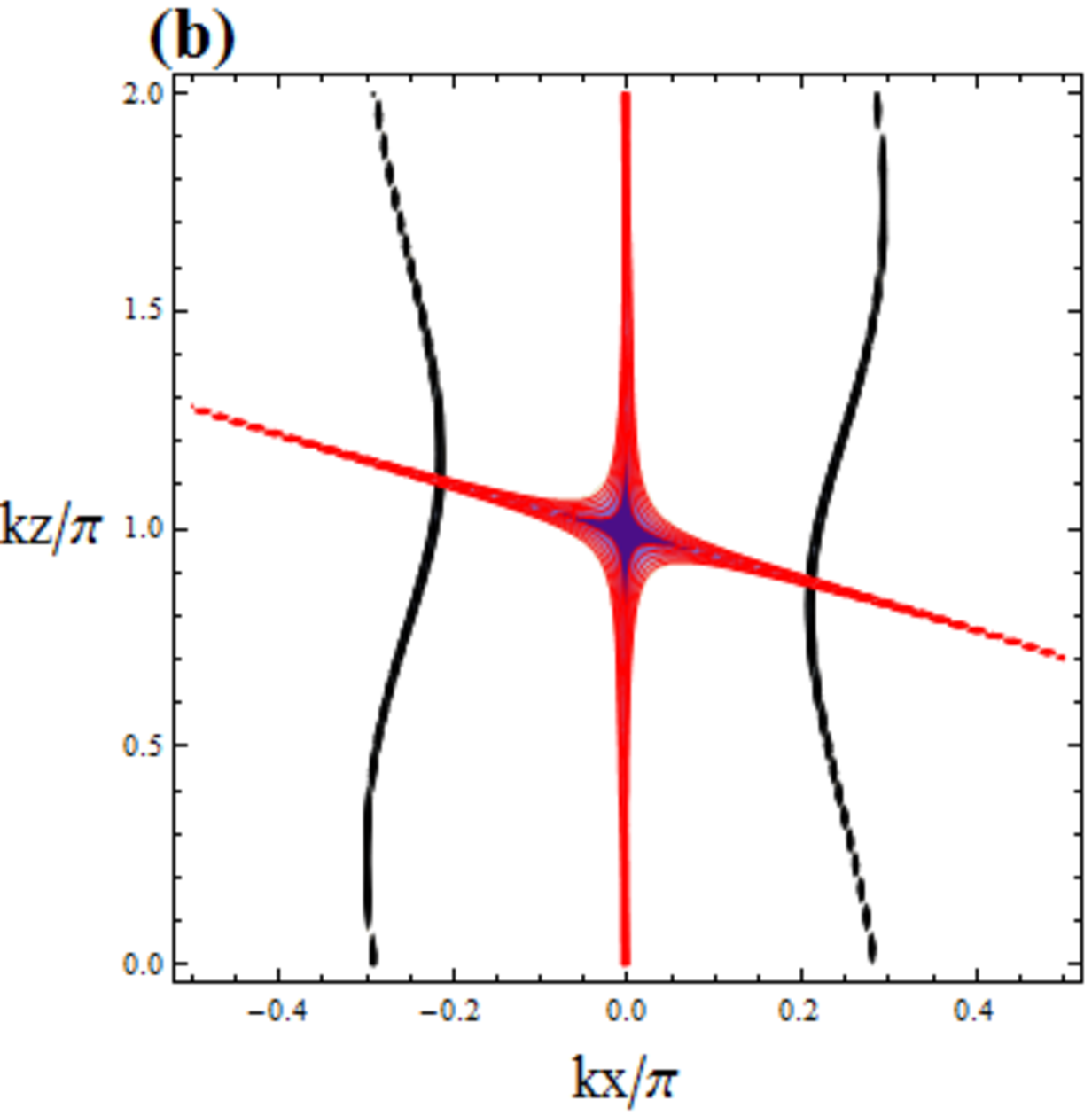}  \\
\includegraphics[width=6cm]{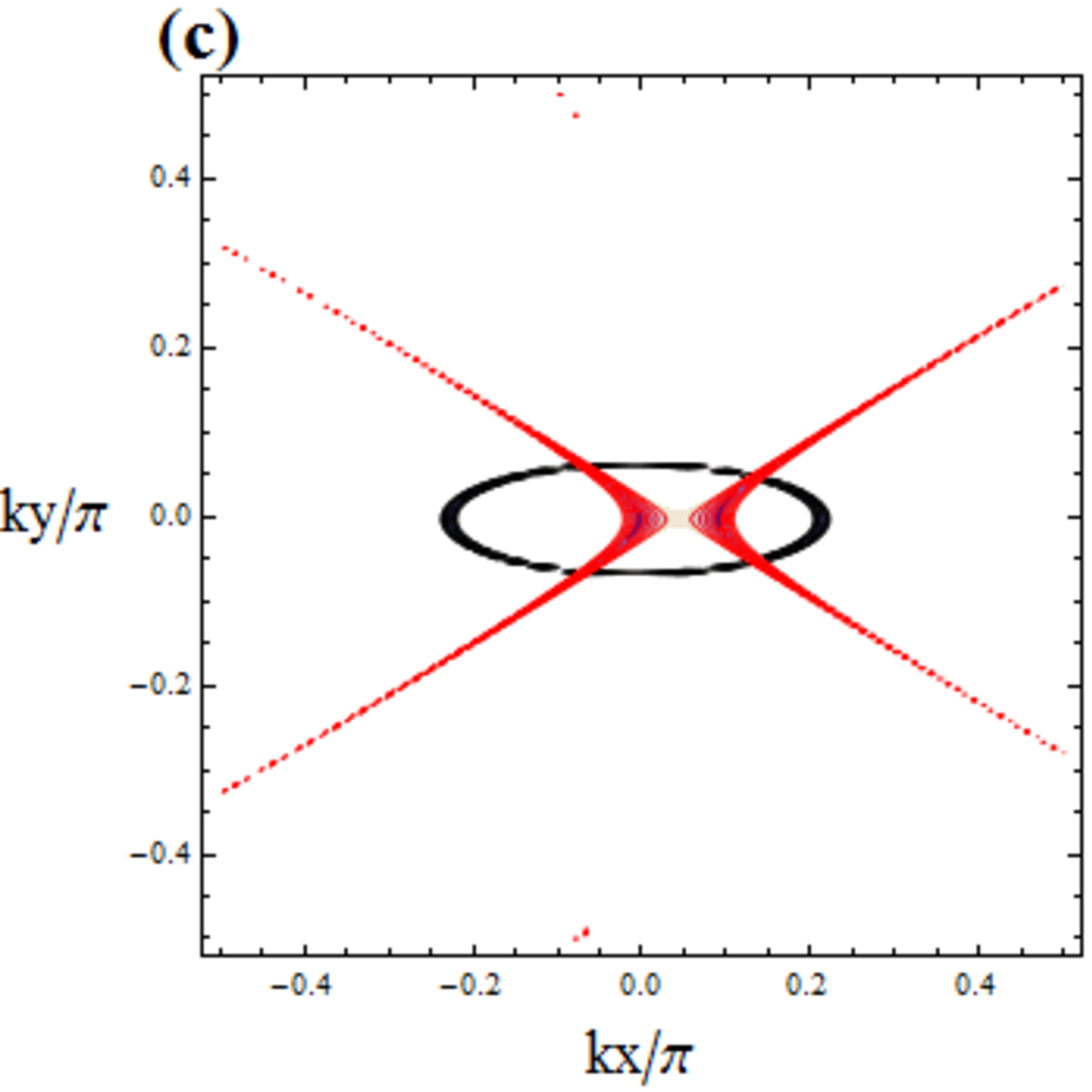}  
\includegraphics[width=6cm]{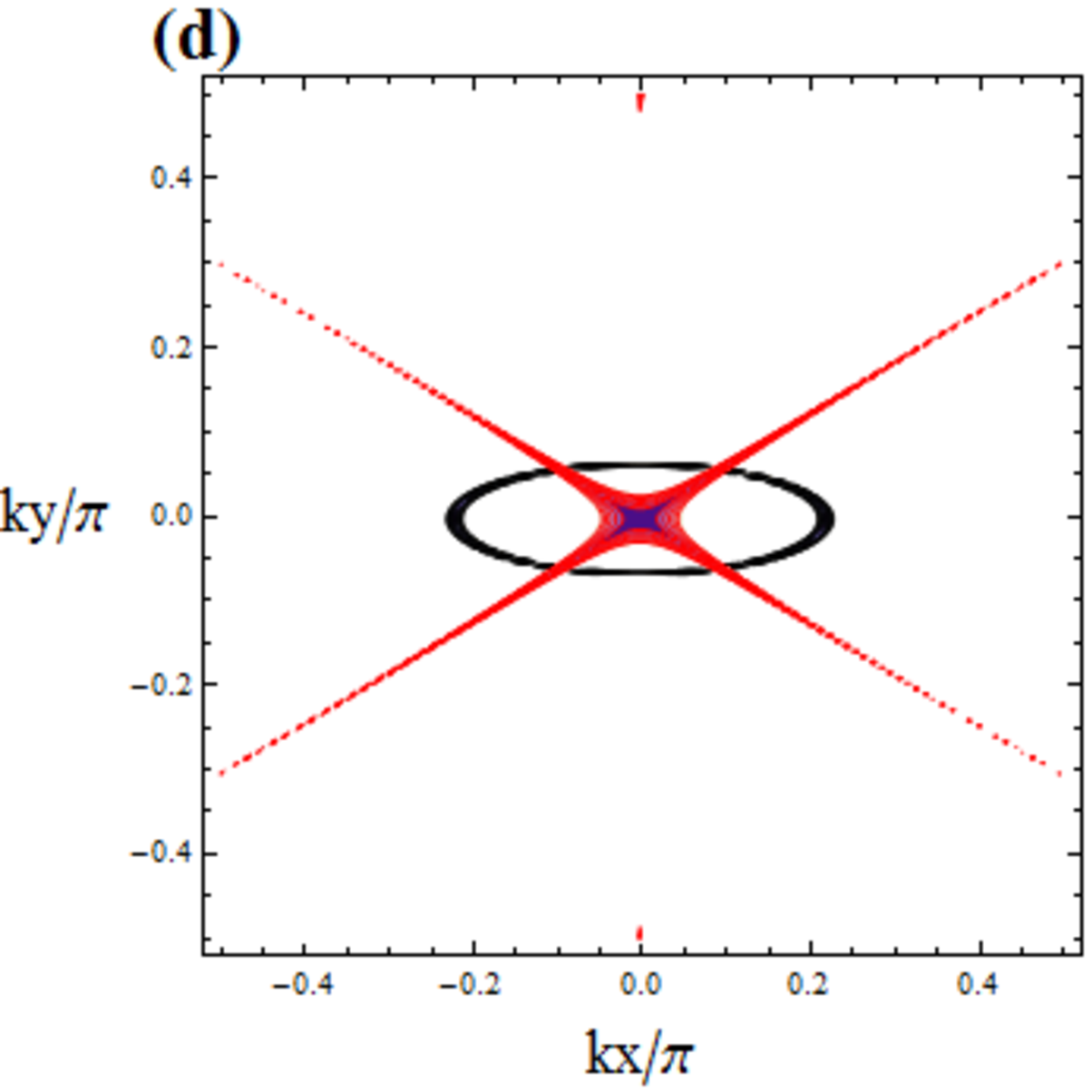}  
\caption{(Color online)
  Fermi line (   black curve) 
 and  nodal line (  red line)
 for $k_z=0$ (a), $k_y = 0$ (b), $k_z/\pi=0.95$ (c), and $k_z/\pi=1.0$ (d), 
 which are calculated from $h_{11}(\bk)=h_{22}(\bk)$ and $h_{12}(\bk)=0$ 
 in Eq.~(\ref{eq:Heff}).  
The intersection of these lines gives the Dirac point. 
}
\label{fig5}
\end{figure}

Here, 
  an effective Hamiltonian   
  for  the bands $E_4(\bk)$ and $E_5(\bk)$ is briefly discussed 
 to comprehend the behavior  of the Dirac point. 
 Noting that  $P_{E4}(\Gamma) = -$ and  $P_{E5}(\Gamma) = + $,  
 we apply 
the method used for the case of \ET.\cite{Suzumura_JPSJ_2016}  
 After replacing $\hat{H}(\bk)$ of Eq.~(\ref{eq:H}) 
 by   $\tilde{H}(\bk)( = \hat{P}(\bk)^{1/2} \hat{H}(\bk)\hat{P}(\bk)^{-1/2})$
 to obtain the real matrix elements,
 $\tilde{H}(\bm{k})$ is rewritten as  
$\hat{H}(\bm{k}) = \tilde{H}^{\rm HH}(\bm{k})+\tilde{H}^{\rm LL}(\bm{k})
  + \tilde{H}^{\rm HL}(\bm{k}),
$
 where the  matrix elements of 
 $\tilde{H}^{\rm HH}(\bm{k})$,  $\tilde{H}^{\rm LL}(\bm{k})$, and 
 $ \tilde{H}^{\rm HL}(\bm{k})$ 
 are expressed  in terms of  the transfer energies of  HOMO-HOMO, LUMO-LUMO, and HOMO-LUMO, respectively.
 Note  that 
 $\tilde{H}^{\rm HH}(\bk) = \tilde{H}^{\rm HH}(-\bk)$ and  
 $\tilde{H}^{\rm LL}(\bk) = \tilde{H}^{\rm LL}(-\bk)$, 
  and that the relation $\tilde{H}^{\rm HL}(\bm{k}) = - \tilde{H}^{\rm HL}(-\bm{k})$ comes from the difference in the symmetry 
 between  the LUMO and HOMO.
 Defining   $|A>$  as the wave function for the maximum eigenvalue of 
    $\tilde{H}^{\rm HH}(\bk)$ and 
  $|B>$  as that for the minimum  eigenvalue of 
    $\tilde{H}^{\rm LL}(\bk)$, 
the 2 $\times$ 2  effective Hamiltonian 
is given by 
\begin{eqnarray}
{H}_{\rm eff}(\bm{k})
 &=& 
\begin{pmatrix}
h_{11}(\bk) & h_{12}(\bk)  \\
h_{21}(\bk) & h_{22}(\bk) 
\end{pmatrix} \ , 
\label{eq:Heff}
\end{eqnarray}
 where 
$ h_{11}(\bk)= <A|\tilde{H}^{\rm HH}(\bk)|A>$,    
$ h_{22}(\bk)= <B|\tilde{H}^{\rm LL}(\bk)|B>$,   
 and 
$h_{12}(\bk) = h_{21}^*(\bk)  = <A|\tilde{H}^{\rm HL}(\bm{k}) |B>$.
 The quantity  $h_{12}(\bk)$  is 
 determined by the combined effect  
  of the interlayer and  intralayer couplings.\cite{Kato_JACS}   
Noting that $E_4(\bk)$ and $E_5(\bk)$ are eigenvalues of 
Eq.~(\ref{eq:Heff}), 
the Dirac point ($E_4(\bk) = E_5(\bk)$)
 is obtained from  $h_{11}(\bk)=h_{22}(\bk)$ and $h_{12}(\bk)=0$, 
 which give the Fermi surface and  nodal plane, respectively.
The loop  of the Dirac point  
 is obtained by the intersection of these two planes,   
 which are shown  in Figs.~\ref{fig5}(a)-\ref{fig5}(d)
 on a 2D plane with  a reduced zone. 
Figures \ref{fig5}(a) and \ref{fig5}(b) correspond to 
Figs.~\ref{fig2}(a) and \ref{fig2}(b), respectively.
For $k_z=0$,
 the Fermi line is  an ellipsoid while the nodal line is given by 
$k_x=0$ (Fig.~\ref{fig5}(a)). 
For $k_y=0$,
 the Fermi line is almost parallel to $k_x=0$ with a bottleneck 
 for $k_z \sim \pi$, 
 while there are two  nodal lines, $k_x = 0$ and $k_z - \pi \simeq - 0.6 \; k_x$ 
 (Fig.~\ref{fig5}(b)).
 With increasing $k_z$  the Fermi surface remains almost the same 
 but   the nodal line varies as follows. 
For $0.91< k_z/\pi < 1.09$, there are two kinds of   
 nodal lines as shown in Fig.~\ref{fig5}(c),
 while only the left line  intersects with the Fermi line 
  for  $0 < k_z/\pi < 0.91$.
 In Fig.~\ref{fig5}(d), the  case of  $k_z/\pi = 1$  
 is shown, where the line  $k_y \simeq \pm 0.6 \; k_x$ 
  is  consistent 
 with the parity  $P_{E4}(Z) = -$ and  $P_{E5}(Z) = - $.
These behaviors suggest that the  nodal plane 
 is periodic   with respect to $k_z$ 
 and the overlap between two nodal planes  occurs in the interval region of 
$0.91< k_z/\pi < 1.09$.
It is found that the Dirac points obtained from Figs.~\ref{fig5}(a)-\ref{fig5}(d)
well  reproduce those of Fig.~\ref{fig3}.

\acknowledgements
The authors are grateful to T. Tsumuraya for useful discussions
 in the early stage of the present work.
One of the authors (Y.S.) thanks T. Kariyado for useful comments and 
also  C. Hotta and T. Osada for helpful comments. 
This work was supported by JSPS KAKENHI Grant Numbers JP15H02108, JP26400355, and JP16H06346.

\newpage

\appendix
 \section{Matrix elements of Hamiltonian}
 The matrix elements of Eq.~(\ref{eq:H_m}) are given by  
\begin{eqnarray}
t_{H1,H1} &=& 
2 b_{1H} \cos \kb \; ,   \\ 
t_{H1,H2} &=& a_H(1+\e^{-i(\kabc)})\; ,   \\ 
t_{H1,H3}  &=&  p_H(1+\e^{-i\kb}+\e^{-i\ka}+\e^{-i(\kab)}) \; , \\ 
t_{H1,H4}  &=&  c_H(1+\e^{i\kc})\; , \\ 
t_{H1,L1}  &=&    b_{1HL}(\e^{i\kb}-\e^{-i\kb}) \; ,\\ %
t_{H1,L2}  &=&  0 \; ,\\ %
t_{H1,L3}  &=&  p_{1HL}+ p_{2HL}\e^{-i\kb}-p_{2HL}\e^{-i\ka} 
           -p_{1HL}\e^{-i(\kab)}  \; ,\\ %
t_{H1,L4}  &=& 0 \; ,\\ %
t_{H2,H2}  &=&   2 b_{2H} \cos \kb  \; ,\\ %
t_{H2,H3}  &=&  c_H(1+\e^{i\kc}) \; ,\\ %
t_{H2,H4}  &=&  q_H(\e^{i(k_x+k_z)}+\e^{i(\kabc)}  +\e^{i\kc}+\e^{i(\kbc)})\; ,\\ %
t_{H2,L1}  &=&   a_{HL}(1-\e^{i(\kabc)})\; ,\\ %
t_{H2,L2}  &=&   b_{2HL}(\e^{ik_y}-\e^{-ik_y}) \; ,\\ 
t_{H2,L3}  &=&  c_{HL}(\e^{-ik_y}-\e^{i(\kbc)}) \; ,\\ %
t_{H2,L4}  &=&  q_{1HL}\e^{i(\kac)}+q_{2HL}\e^{i(\kabc)}- q_{2HL}\e^{i\kc}-q_{1HL}\e^{i(\kbc)}\; ,\\ %
t_{H3,H3}  &=&  2 b_{1H} \cos \kb \; ,\\ %
t_{H3,H4}  &=& a_H(\e^{i\kb}+\e^{i(k_x+k_z)}) \; ,\\ %
t_{H3,L1}  &=&   p_{2HL} + p_{1HL}\e^{i\kb} -p_{1HL}\e^{i\ka}-p_{2HL}\e^{i(\kab)} \; ,\\ %
t_{H3,L2}  &=&  0 \; ,\\ %
t_{H3,L3}  &=&  b_{1HL}(\e^{i\kb}-\e^{-i\kb})\; ,\\ %
t_{H3,L4}  &=&  0 \; ,\\ %
t_{H4,H4}  &=&  2 b_{2H} \cos \kb \; ,\\ %
t_{H4,L1}  &=&   c_{HL}(\e^{-i\kb}-\e^{i(\kb-\kc)})\; ,\\ %
t_{H4,L2}  &=&  q_{2HL}\e^{-i(\kac)}+ q_{1HL}\e^{-i(\kabc)}-q_{1HL}\e^{-i\kc} -q_{2HL}\e^{-i(\kbc)}\; ,\\ %
t_{H4,L3}  &=&  a_{HL}(\e^{-i\kb}-\e^{-i(\kac)}) \; ,\\ %
t_{H4,L4}  &=&  b_{2HL}(\e^{i\kb}-\e^{-i\kb}) \; ,\\ %
t_{L1,L1}  &=&  \Delta E + 2 b_{1L} \cos \kb \; ,\\ %
t_{L1,L2}  &=&  a_{L}(1+\e^{-i(\kabc)})\; ,\\ %
t_{L1,L3}  &=&  p_{L}(1+\e^{-i\kb}+\e^{-i\ka}+\e^{-i(\kab)})\; ,\\ %
t_{L1,L4}  &=&  c_{L}(\e^{i\kb}+\e^{i(-\kb + \kc)}) \; ,\\ %
t_{L2,L2}  &=&  \Delta E + 2 b_{2L} \cos \kb \; ,\\ %
t_{L2,L3}  &=&   c_{L}(\e^{-i\kb}+\e^{i(\kbc)})\; ,\\ %
t_{L2,L4}  &=&  q_{L}(\e^{i(\kac)}+ \e^{i(\kabc)}+\e^{i\kc}+\e^{i(\kbc)}) \; ,\\ %
t_{L3,L3}  &=& \Delta E + 2 b_{1L} \cos \kb \; ,\\ %
t_{L3,L4}  &=& a_{L}(\e^{i\kb}+\e^{i(k_x+k_z)})\; ,\\ %
t_{L4,L4}  &=&  \Delta E + 2 b_{2L} \cos \kb \; , %
\label{eq:matrix_element} 
\end{eqnarray}
 where $t_{\beta,\alpha} = t_{\alpha, \beta}^*$.


\end{document}